\documentclass[preprint,12pt]{elsarticle}

\usepackage[margin=1in]{geometry}
\usepackage{amsmath}

\usepackage{graphicx} 
\usepackage{subcaption}

\usepackage{changes}
\usepackage{comment}
\usepackage{hyperref}

\usepackage{xr-hyper}
\makeatletter
\newcommand*{\addFileDependency}[1]{% argument=file name and extension
  \typeout{(#1)}
  \@addtofilelist{#1}
  \IfFileExists{#1}{}{\typeout{No file #1.}}
}
\makeatother

\usepackage[utf8]{inputenc}
\usepackage[T1]{fontenc}
\newcommand{\smallotimes}{\mathbin{\mathpalette\make@small\otimes}}

\date{}

\begin{document}

\begin{frontmatter}


% \title{Embedding Aided Interlayer Similarity 
% %Reveals
% the Robustness and Reducibility of Multiplex Network }

% \title{Quantifying the Robustness and Reducibility of Multiplex Network with
% Embedding Aided Interlayer Similarity }

\title{Assessing the Robustness and Reducibility of Multiplex Networks with Embedding-Aided Interlayer Similarities}
% \tnoteref{t1}}
% \tnotetext[t1]{Haoran Nan and Senquan Wang contributed equally to this work.}

\author[1]{Haoran Nan\fnref{fn1}}
\author[1]{Senquan Wang\fnref{fn1}}
\author[1]{Chun Ouyang}
\author[1]{Yanchen Zhou}
\author[1]{Weiwei Gu\corref{cor1}}

\cortext[cor1]{Corresponding author}
\fntext[fn1]{These authors contributed equally to this work.}

\affiliation[1]{organization={College of Information Science and Technology, Beijing University of Chemical Technology},
                % addressline={Address line}, % Add specific address if needed
                city={Beijing},
                % postcode={Postal Code}, % Add postal code if known
                % state={State}, % Add state if applicable
                country={China}}

\ead{weiweigu@mail.buct.edu.cn}  

\begin{abstract}
The study of interlayer similarity of multiplex networks helps to understand the intrinsic structure of complex systems, revealing how changes in one layer can propagate and affect others, thus enabling broad implications for transportation, social, and biological systems. Existing algorithms that measure similarity between network layers typically encode only partial information, which limits their effectiveness in capturing the full complexity inherent in multiplex networks. To address this limitation, we propose a novel interlayer similarity measuring approach named \textbf{E}mbedding \textbf{A}ided in\textbf{T}erlayer \textbf{Sim}ilarity (EATSim). EATSim concurrently incorporates intralayer structural similarity and cross-layer anchor node alignment consistency, providing a more comprehensive framework for analyzing interconnected systems. Extensive experiments on both synthetic and real-world networks demonstrate that EATSim effectively captures the underlying geometric similarities between interconnected networks, significantly improving the accuracy of interlayer similarity measurement. Moreover, EATSim achieves state-of-the-art performance in two downstream applications: predicting network robustness and network reducibility, showing its great potential in enhancing the understanding and management of complex systems.
\end{abstract}

%% Keywords

\end{frontmatter}

\begin{center} %arXiv版本，说明APS版权信息
    \vspace{3em}
    \footnotesize
    \textbf{This manuscript has been accepted for publication in \textit{Physical Review E}.\\
    © 2025 American Physical Society. This version may differ from the final published version. The final version will be available at \url{https://journals.aps.org/pre}}
\end{center}
\vspace{3em}

% \begin{document}

\flushbottom
\thispagestyle{empty}

\section{Introduction}
Many complex systems are characterized by diverse and interdependent interactions. A comprehensive understanding of these systems requires analyzing their interconnected nature rather than treating each interaction in isolation~\cite{Kivelä2014Multilayer,DeDomenico2023More}. Multilayer networks provide a robust framework to capture the complexity of these interdependencies, enabling integrated analysis of both natural and artificial systems \cite{DeDomenico2023More,Gao2012Networks,Martinez2017Survey}. Recent research underscores that analyzing complex systems through the multilayer network framework not only advances scientific discovery \cite{DeDomenico2013Mathematical,Zeng2017science,Wang2023Scientific} but also enhances the accuracy of various network-based applications, such as node classification \cite{wang2019heterogeneous} and link prediction \cite{gu2023improving,gu2024pay}.

Understanding interlayer relationships in multilayer networks is crucial for analyzing and optimizing complex systems across infrastructure, cybersecurity, and biological domains \cite{DeDomenico2023More,DeDomenico2015Structural,Kleineberg2017Geometric,Schieber2017Quantification,Xu2023Interconnectedness,Wang2023A}. Structurally similar layers exhibit stronger interdependence, enhancing information flow and system stability, while dissimilar layers often lead to weaker interactions and reduced resilience \cite{Ghavasieh2020Enhancing}. Structural similarity metrics provide a quantitative framework for assessing interlayer dependencies, aiding in network optimization, failure prediction, and anomaly detection~\cite{Kleineberg2017Geometric,Schieber2017Quantification,Najari2019Link}. A deeper understanding of these relationships can facilitate improvements in real-world applications, from transportation efficiency to cybersecurity risk assessment. 

Existing methods for measuring interlayer similarity in multiplex networks are often designed for specific systems, focusing either on local structural properties or global degree distribution. For example, degree distribution-based techniques \cite{Najari2019Link,Zhang2020Measuring,Mohapatra2022Core} quantify similarity by comparing node degree distributions across layers, prioritizing high-degree nodes while neglecting the role of low-degree nodes. Community structure-based approaches \cite{Ghawi2022A,Faqeeh2018Characterizing,Calderone2016Comparing} assess interlayer similarity by aligning communities and examining clustering within them. While these methods address some limitations of degree-based metrics, their high computational complexity and limited scalability remain significant challenges, especially for large-scale multilayer networks \cite{Kim2015Community}. To capture higher-order similarities, macroscopic approaches have been proposed, utilizing information theories such as Von Neumann entropy \cite{DeDomenico2015Structural}, Jensen–Shannon divergence \cite{Schieber2017Quantification}, and mutual information~\cite{Kleineberg2017Geometric,Felippe2024Network}. However, these methods primarily rely on global statistical properties and often fail to capture fine-grained structural differences at the node or local level, limiting their ability to represent the intricate and multifaceted nature of multiplex networks.

In recent years, network embedding, also known as network representation learning, has garnered significant attention, leading to substantial advancements in various fields of network science. Network embedding algorithms encode network structures into continuous vector representations while preserving the structural and relational properties of networks \cite{Hamilton2017Representaion,Cui2019A,Gu2025MWTP}. These algorithms have greatly facilitated tasks such as node classification \cite{gu2021discovering}, link prediction \cite{gu2024pay}, and node clustering \cite{gu2021principled}, demonstrating remarkable potential for solving various graph-related problems \cite{Xiao2022Graph,kallestad2023general}. In this paper, we leverage network embedding and propose an interlayer relation quantification algorithm, termed \textbf{E}mbedding \textbf{A}ided in\textbf{T}erlayer \textbf{Sim}ilarity (EATSim). EATSim quantifies interlayer similarity by considering both intralayer topological similarity and cross-layer alignment consistency. This is achieved by combining the Pairwise Euclidean Distance (PED) loss, which computes distances between node embeddings in vector space, with the Aligned Euclidean Distance (AED) loss, which assesses cross-layer alignment after orthogonal transformation.

The idea of measuring intralayer local topological similarity using network embedding was introduced by Gu et al. \cite{gu2021principled} to identify the optimal embedding dimensions by calculating the cosine similarity of the embedding vector between nodes. Zhang et al. \cite{gu2017hidden} studied the hidden geometric property of network embedding by computing the Euclidean distances between embedding vectors of nodes. Inspired by these works, Pairwise Euclidean Distance (PED) loss measures intralayer Euclidean distances between the embedding vectors of all node pairs and computes the absolute differences between these distances to quantify topological dissimilarity across different layers. 

In addition to intralayer local topological similarity, cross-layer alignment consistency, which reflects global structural similarity, is also essential for measuring network similarity. Mikolov et al. \cite{Mikolov2013Exploiting} explored language similarity by aligning word vector spaces through linear transformations, ensuring that semantically similar words across languages were represented by comparable vectors. Inspired by this method, Aligned Euclidean Distance (AED) loss applies a geometric orthogonal transformation to node embedding vectors across layers. By using cross-layer connected nodes as alignment anchors, AED loss computes the alignment error to quantify the differences between network layers.

EATSim integrates both PED loss and AED loss, and we validate its effectiveness through two tasks: interconnected network robustness prediction and network reducibility measurement. Multiplex networks are inherently fragile and highly susceptible to disturbances~\cite{Kleineberg2017Geometric,Faqeeh2018Characterizing}; even minor disturbances in a single layer can lead to large-scale system failures through the failure propagation mechanism between layers~\cite{Buldyrev2010Catastrophic}. In this study, we demonstrate that EATSim accurately predicts the robustness of interconnected networks, highlighting the critical role of interlayer dependencies in enhancing network resilience. This insight provides a theoretical foundation for the design and construction of robust complex systems, including infrastructure and supply chain networks.

In the era of big data, the rapid expansion of multilayer networks presents challenges in computation, storage, and information extraction. Multilayer network reduction addresses these issues by reducing redundancy, preserving key topological structures, and improving computational efficiency. This approach enhances data analysis across fields such as artificial intelligence, intelligent transportation, and social network analysis \cite{Battiston2014Structural}. A key strategy is to simplify multiplex structures by minimizing the number of layers while retaining essential information \cite{DeDomenico2015Structural}. EATSim tackles this challenge by identifying an optimal layer subset through relation aggregation based on layer similarity. Experiments on synthetic and real-world networks demonstrate its effectiveness in network reducibility measurement task.

\section{Methodology}

\subsection{Network representation with node2vec}
We consider a multiplex network consisting of the same set of $N$ nodes, where each layer represents a distinct type of relation and nodes across different layers have one-to-one interlinks or interactions. The multiplex network is denoted as $G=\{G^{(\alpha)}\}$, where $\alpha=1,2,...,L$, and $L$ represents the number of relations. In each layer $\alpha$, $G^{(\alpha)} = (V^{(\alpha)}, E^{(\alpha)})$, with $V^{(\alpha)}$ representing the set of nodes and $E^{(\alpha)}$ representing different edge relations. To effectively capture the structure and properties of multiplex networks, we apply network embedding techniques, which encode the graph structure into low-dimensional vector representations while preserving local connectivity patterns as well as global network structures~\cite{Cui2019A}. Network embedding has been extensively studied, with various algorithms ranging from shallow embedding algorithms such as node2vec~\cite{Grover2016node2vec} and LINE~\cite{tang2015line} to deep neural networks such as GraphSage~\cite{hamilton2017inductive} and GCNs~\cite{kipf2016semi}. 

In this study, we find that even shallow linear neural network representation algorithms, such as node2vec \cite{Grover2016node2vec} can quantify interlayer similarity, and support the prediction of network robustness and reducibility. Node2Vec utilizes a flexible random walk strategy in combination with the skip-gram model \cite{mikolov2013efficient}, which was originally developed for natural language processing to learn word embeddings and has been adapted for network representation learning. This approach embeds nodes into a continuous vector space, preserving both local and global network structures. By leveraging these embeddings, node2vec enhances the ability to capture higher-order proximities, leading to improved performance in tasks such as link prediction~\cite{gu2023improving} and community detection~\cite{kojaku2024network} in complex networks.

Since the networks analyzed in this paper are not very large, we set the embedding dimension $d$ to 32 based on the approach proposed in~\cite{gu2021principled}. 
Additionally, the number of walks and the window size are both set to 10, consistent with the typical values used in the original node2vec implementation~\cite{Grover2016node2vec}. For the walk length, we refer to the settings adopted in~\cite{gu2021principled} and~\cite{gu2017hidden}, and set it to 10 accordingly. We also set the parameters $p$ and $q$ to 1, as is common practice.
The final embedding matrix for layer $\alpha$ is denoted as $\mathbf{X}^{(\alpha)}$, where each row represents a node’s embedding vector, $\mathbf{x}^{(\alpha)}_i$. Layer similarity in the multiplex network is measured by quantifying both intralayer topological discrepancies and cross-layer alignment inconsistencies, based on the learned node embeddings.
%the discrepancy between simulations and observations

\subsection{Pairwise Euclidean Distance (PED) loss}

To quantify the intralayer topological difference, we use the hidden geometric space of the network embeddings~\cite{gu2017hidden}, transforming the layer similarity problem into a distance comparison task. Specifically, we calculate the Euclidean distance between node pairs within layer $\alpha$ by computing the $L_2$ norm in a $d$-dimensional Euclidean space, as defined in Eq.~(\ref{eq:Euclidean}):
\begin{equation}
   \centering
    \label{eq:Euclidean}
   E_{dis}(\mathbf{x}^{(\alpha)}_{i},\mathbf{x}^{(\alpha)}_{j}) =  {\sqrt{\sum_{k=1}^{d} (x_{i k}^{(\alpha)}-x_{j k}^{(\alpha)})^{2}}}.
\end{equation}

The intralayer topological difference is then determined by the average accumulated absolute Euclidean distance between node pairs. Eq.~(\ref{eq:L_PED}) computes the Pairwise Euclidean Distance (PED) loss between two layers, serving as a comprehensive metric that quantifies the local structural discrepancies across layers.

\begin{equation}
    \centering
    \label{eq:L_PED}
    \mathcal L_{\mathrm{PED}}(\mathbf {X}^{(\alpha)}, \mathbf{X}^{(\beta)}) =  \frac{2}{N(N-1)}   \sum_{i<j}  \left|  E_{dis}(\mathbf{x}^{(\alpha)}_{i},\mathbf{x}^{(\alpha)}_{j}) - E_{dis}(\mathbf{x}^{(\beta)}_{i},\mathbf{x}^{(\beta)}_{j}) \right|,
\end{equation}
where $\mathbf{X}^{(\alpha)}$ and $\mathbf{X}^{(\beta)}$ represent the matrices of the embedding from layers $\alpha$ and $\beta$, respectively.

PED loss operates under the assumption that layers with similar structures exhibit comparable local topological properties. To maintain embedding consistency, we fix the random seed during the representation learning process. This removes stochastic variations, ensuring that for two networks with identical structures, the PED loss is zero, reflecting a high degree of structural alignment.

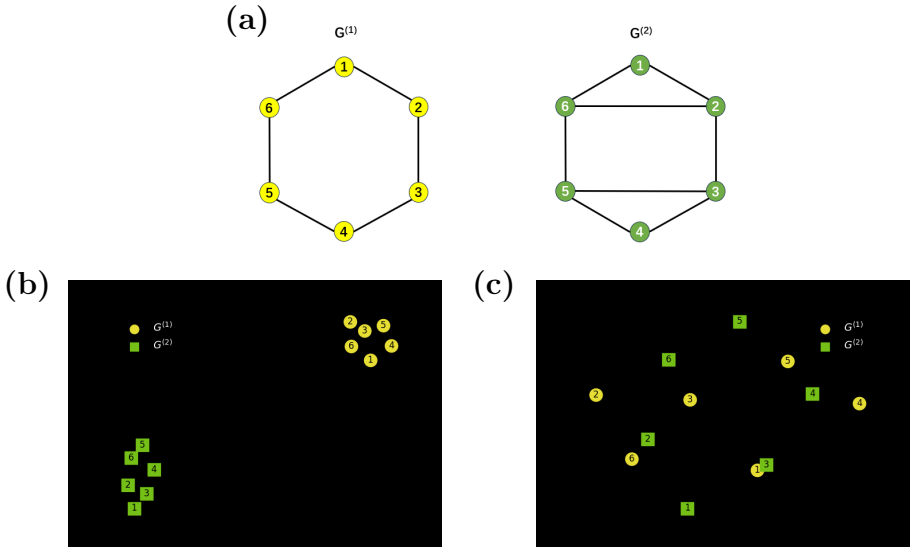
\begin{figure}[htbp]
    \centering
    % First image at the top center
    \begin{subfigure}[b]{0.4\textwidth}
        \centering
        \includegraphics[width=\textwidth]{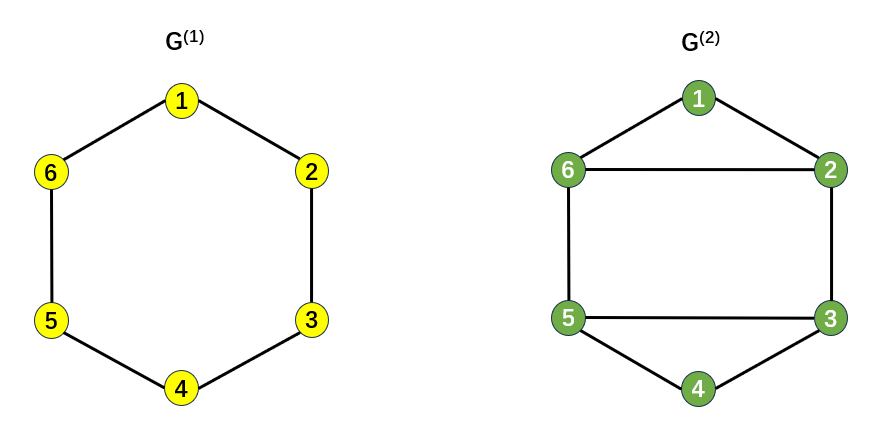}
        % \caption{}
        \label{fig:image1}
        \begin{picture}(0,0)
            \put(-100,100){\textbf{(a)}}
        \end{picture}
    \end{subfigure}
    \\[-0.5em]  % 强制换行并略微收紧行间距
    % Two images in the bottom row
    \begin{subfigure}[b]{0.3\textwidth}
        \centering
        \includegraphics[width=\textwidth]{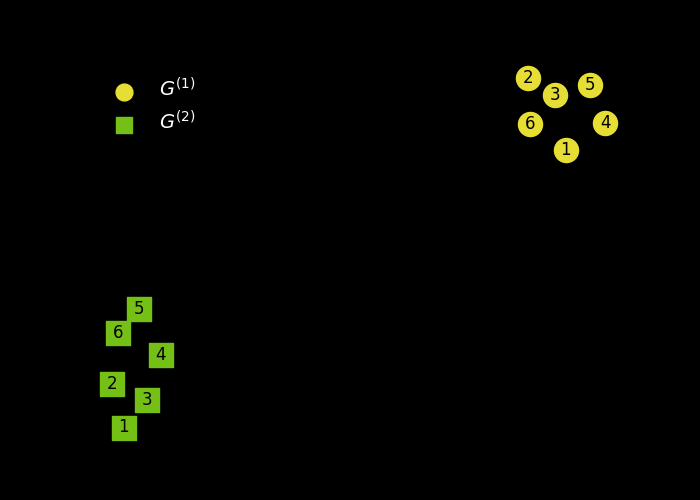}
        % \caption{}
        \label{fig:image2}
        \begin{picture}(0,0)
            \put(-95,110){\textbf{(b)}}
        \end{picture}
    \end{subfigure}
    \hspace{1cm}
    \begin{subfigure}[b]{0.3\textwidth}
        \centering
        \includegraphics[width=\textwidth]{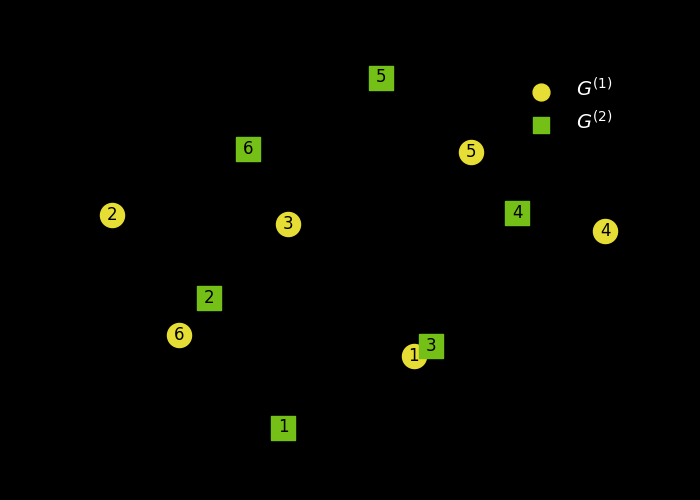} 
        % \caption{}
        \label{fig:image3}
        \begin{picture}(0,0)
            \put(-95,110){\textbf{(c)}}
        \end{picture}
    \end{subfigure}
       \caption{\textbf{A toy example illustrating the effectiveness of PED and AED.} \textbf{(a)} Diagram of two networks with the same number of nodes but different topologies. \textbf{(b)} Visualization of network embeddings after dimensionality reduction to 2 using the Uniform Manifold Approximation and Projection (UMAP) method. \textbf{(c)} Visualization of network embeddings after aligning anchor nodes using an orthogonal transformation.}
    \label{fig:case}
\end{figure}

To illustrate the effectiveness of the proposed PED loss, we present two simple toy networks, as shown in Fig.~\ref{fig:case}\textbf{(a)}. Both networks, $G^{(1)}$ and $G^{(2)}$, have the same set of nodes but differ in their connectivity patterns. Specifically, $G^{(2)}$ includes two additional edges---(2, 6) and (3, 5)---compared to $G^{(1)}$.
%This structural variation influences the node embeddings, as depicted in Fig.~\ref{fig:case}\textbf{(b)}, where we visualize the low-dimensional embeddings of $G^{(1)}$ and $G^{(2)}$ obtained through UMAP \cite{McInnes2018Umap}. The embeddings of $G^{(1)}$ exhibit a more symmetric distribution, likely due to its homogeneous local connectivity patterns being preserved in biased random walks of node2vec. 
This structural difference leads to noticeable variations in the resulting node embeddings, as illustrated in Fig.~\ref{fig:case}\textbf{(b)}. The figure shows the low-dimensional embeddings of $G^{(1)}$ and $G^{(2)}$ obtained using UMAP \cite{McInnes2018Umap}. Notably, the embeddings of $G^{(1)}$ exhibit a more symmetric spatial distribution, which can be attributed to the preservation of its relatively homogeneous local connectivity patterns during the biased random walks employed by node2vec.

In contrast, the embeddings of $G^{(2)}$ exhibit clear community structures, which may be attributed to the disruption of the original local connectivity patterns caused by the introduction of long-range edges. The proposed PED loss effectively captures these spatial variations by measuring local structural discrepancies based on the relative positions of the nodes in the embedding space.

\subsection{Aligned Euclidean Distance (AED) loss}
Cross-layer alignment similarity, first introduced by Mikolov et al.~\cite{Mikolov2013Exploiting} in natural language processing, is applied to quantify the structural similarity between layers. Inspired by this, we propose the Aligned Euclidean Distance (AED) loss to measure global dissimilarity by assessing cross-layer alignment error. AED applies an orthogonal transformation to rotate and translate embedding representations from different layers into a unified space as formulated in Eq.~(\ref{eq:align_2}), to ensure the scale consistency between embeddings. 

\begin{equation}
    \centering
    \label{eq:align_2}
    \mathbf{W}^{\mathrm{o p t}}=\operatorname{a r g} \underbrace{\operatorname* {m i n}}_{\{\mathbf{W} \left| \mathbf{W}^{-1}=\mathbf{W}^{T} \right\}} \| \mathbf{ X^{(\alpha)}} \mathbf{W}-\mathbf{ X^{(\beta)}} \|_{F}^{2}=\operatorname* {a r g} \underbrace{\operatorname* {m a x}}_{\{\mathbf{W} \left| \mathbf{W}^{-1}=\mathbf{W}^{T} \right\}} \mathrm{T r} \left[ \mathbf{W}^{T} \mathbf{ X^{(\alpha)}}^{T} \mathbf{ X^{(\beta)}} \right] = \tilde{{\bf U}} \tilde{{\bf V}}^{T},
\end{equation}
where $\mathbf{W}^{\mathrm{o p t}}$ is the optimal transformation matrix to minimize the alignment difference between $\mathbf{ X^{(\alpha)}}\mathbf{W}$ and $\mathbf{ X^{(\beta)}}$, obtained by Singular Value Decomposition (SVD) of the matrices $\mathbf{X}^{(\alpha)}\!^{T} \mathbf{X}^{(\beta)} = \tilde{\bf U} \tilde{\Sigma} \tilde{\bf V}^{T}$ and $\tilde{\Sigma}$ is a singular value matrix, where the values are arranged in decreasing order.

Nodes connected by cross-layer links serve as alignment anchors, and we compute the absolute Euclidean distances between anchor nodes to quantify the global topological differences between layers, see Eq.~(\ref{eq:L_AED}):

\begin{equation}
    % \centering
    \label{eq:L_AED}
    \mathcal L_{\mathrm{AED}}(\mathbf {X}^{(\alpha) \prime}, \mathbf{X}^{(\beta)})=  \frac{1}{N}   \sum_{k=0}^{N} 
    \| \mathbf{x}^{(\alpha) \prime}_k, \mathbf{x}^{(\beta)}_k \|_{2},
\end{equation}
where $\mathbf {X}^{(\alpha) \prime}=\mathbf{ X^{(\alpha)}} \mathbf{W}^{\mathrm{o p t}}$ and $\bf X^{(\beta)}$ are the aligned embedding matrices.

The node alignment process improves layer evaluation, with a smaller AED loss indicating greater interlayer similarity. Similar to the PED loss, we fix the random seed in our experiments to produce consistent embeddings, ensuring that structurally identical layers yield an AED loss of zero.

As shown in Fig.~\ref{fig:case}\textbf{(b)}, the PED loss captures local structural differences by measuring node positioning in the embedding space. However, directly comparing the embeddings of $G^{(1)}$ and $G^{(2)}$ for global structural differences is not feasible, as their embedding spaces are not the same. To overcome the space limitation, we introduce the AED loss, which employs anchor nodes to map node embeddings into a shared space. As illustrated in Fig.~\ref{fig:case}\textbf{(c)}, this alignment mitigates the impact of spatial inconsistency, providing a more reliable measure of global structural differences.

\subsection{Measuring interlayer similarity with EATSim}
The final network similarity metric, EATSim, integrates the intralayer PED loss and the cross-layer AED loss. To balance their relative importance, we introduce a hyperparameter \(\omega\) in Eq.~(\ref{eq:D}). The term $D(G^{(\alpha)}, G^{(\beta)})$ quantifies the overall dissimilarity between layers $\alpha$ and $\beta$, where a larger value indicates greater dissimilarity between the two networks.

\begin{equation}
    \label{eq:D}
    D(G^{(\alpha)},G^{(\beta)}) = \omega \mathcal L_{\mathrm{PED}} + (1-\omega) \mathcal L_{\mathrm{AED}}.
\end{equation}

A comparative analysis of $\omega$ values ranging from 0 to 1 (see Supplementary Fig.~1 for details~\cite{SI}) reveals a non-monotonic relationship between $\omega$ and the Pearson correlation to the network robustness indicator, with the maximum correlation observed at $\omega = 0.5$. Accordingly, we Iet $\omega = 0.5$ in this study to balance the contributions of PED loss and AED loss. The resulting EATSim metric, which jointly captures intralayer and cross-layer similarities between networks $G^{(\alpha)}$ and $G^{(\beta)}$, is defined in Eq.~(\ref{eq:EATSim}):

\begin{equation}
    % \centering
        \label{eq:EATSim}
    EATSim(G^{(\alpha)}, G^{(\beta)}) = 1- D(G^{(\alpha)}, G^{(\beta)}).
\end{equation}

In the following section, we validate the effectiveness of EATSim on both synthetic and real-world networks, focusing on two key scenarios: network robustness prediction and network reducibility measurement.

\section{Experiments}
\subsection{Validating the effectiveness of EATSim on synthetic multiplex networks}
\label{sec:3_1}

Quantifying similarity in multiplex networks remains a challenging task. Existing network similarity algorithms either rely on specific assumptions or are tailored to particular network structures, making it difficult to generalize across diverse types of networks. To validate the effectiveness of EATSim, we begin by adjusting interlayer similarity via rewiring the connections of nodes in Barabási–Albert (BA) networks. The rewiring probabilities are applied to control the topological similarity in the generated networks. Additionally, we employ the Geometric Multiplex Model (GMM)~\cite{Kleineberg2016Hidden} to generate synthetic multiplex networks with adjustable angular and radial interlayer correlations. Then, we use the rewiring probabilities of BA networks and the interlayer geometric correlations in GMM networks as benchmarks, examining the relationships between these benchmarks and EATSim.
%-
Edge overlap is a direct indicator of quantifying interlayer similarity in interconnected networks. A high edge overlap suggests that node relationships are similar across layers, pointing to correlated interactions or common patterns of behavior~\cite{DeDomenico2015Structural}. Based on this premise, we first generate a BA network with a network size of \(N = 1,000\) and a degree distribution \(P(k) \sim k^{-3}\). We then generate 19 similar BA networks by varying the rewiring probabilities for the edges of the original network, with probabilities ranging from 5\% to 95\%. This rewiring strategy allows us to generate pairs of interconnected networks with controllable interlayer topological similarities. As the rewiring probability increases, the interlayer edge overlap decreases, leading to fewer shared structures and, consequently, lower interlayer similarity. 

\begin{figure}[htbp]
    \centering
    \begin{subfigure}[b]{0.42\textwidth}
        \centering
        \includegraphics[width=\textwidth]{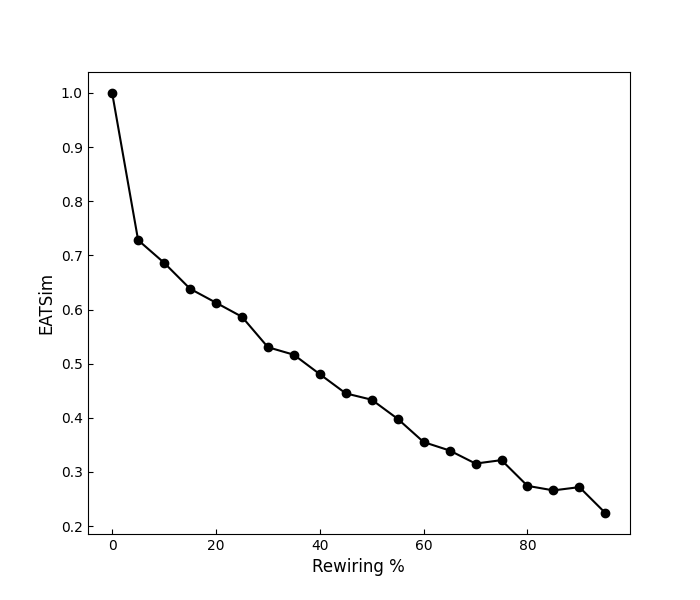}
        % \caption{}
        \label{fig:rewire_}
        \begin{picture}(0,0)
            \put(-110,180){\large \textbf{(a)}}
        \end{picture}
    \end{subfigure}
    \hspace{0.5cm} % 调整这个值来设置间距
    \begin{subfigure}[b]{0.42\textwidth}
        \centering
        \includegraphics[width=\textwidth]{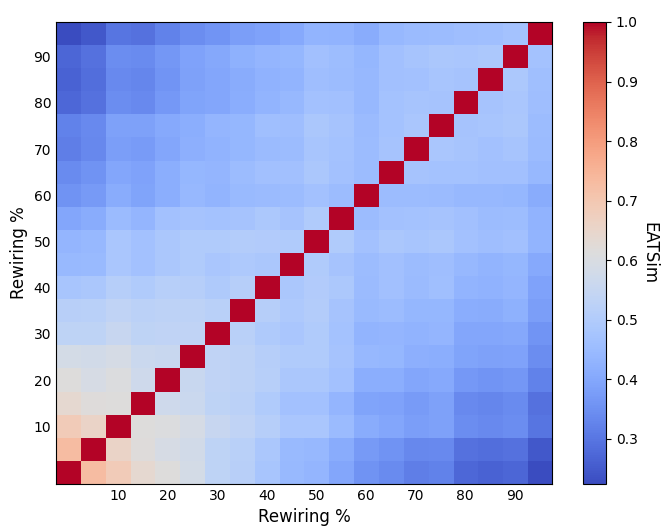}
        % \caption{}
        \label{fig:rewire_a}
        \begin{picture}(0,0)
            \put(-110,180){\large \textbf{(b)}}
        \end{picture}
    \end{subfigure}

    \vspace{0.05cm}

    \begin{subfigure}[b]{0.45\textwidth}
        \centering
        \includegraphics[width=\textwidth]{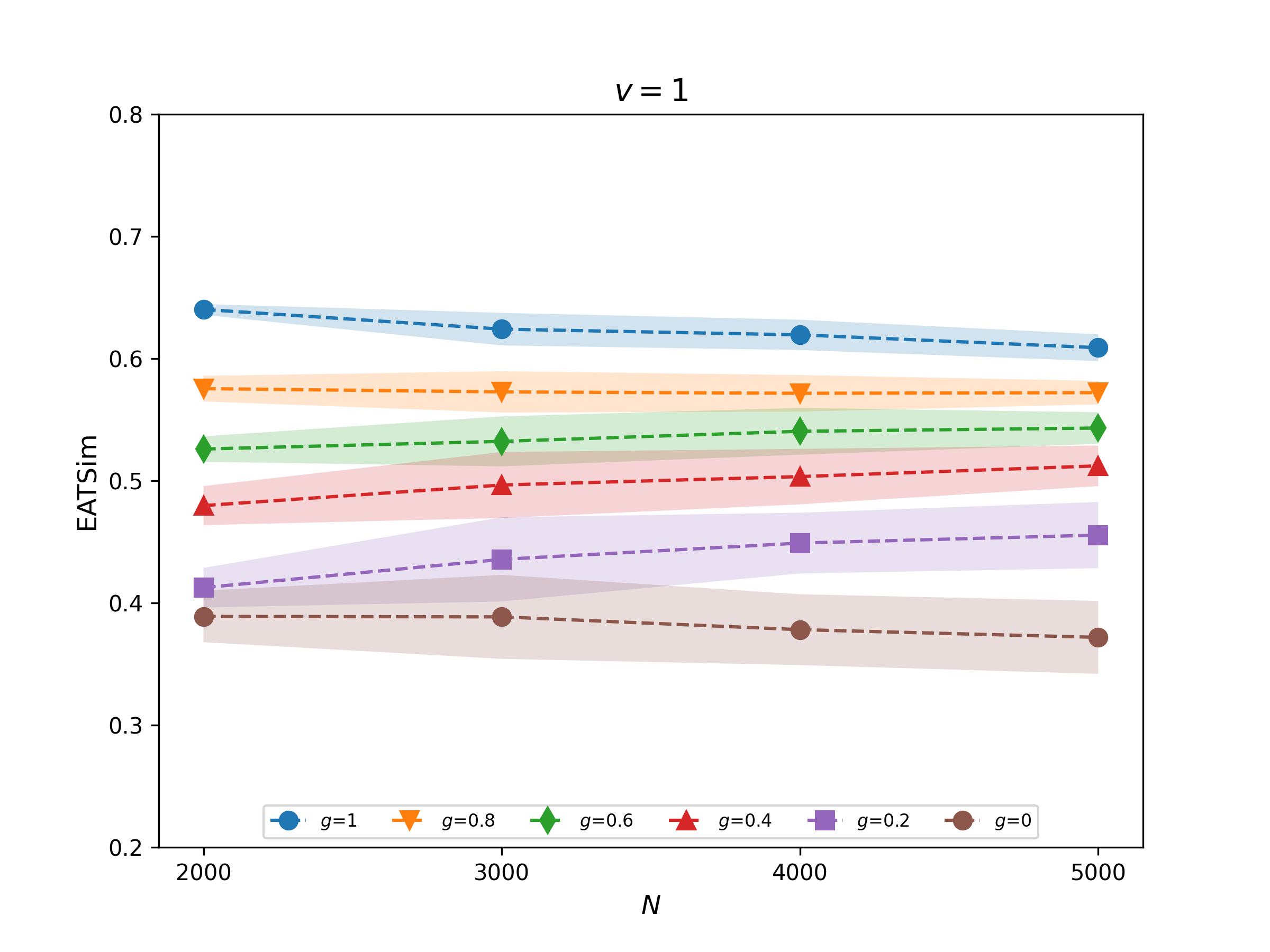}
        % \caption{}
        \label{fig:EATSim_g}
        \begin{picture}(0,0)
            \put(-110,160){\large \textbf{(c)}}
        \end{picture}
    \end{subfigure}
    \hspace{0.1cm} % 调整这个值来设置间距
    \begin{subfigure}[b]{0.45\textwidth}
        \centering
        \includegraphics[width=\textwidth]{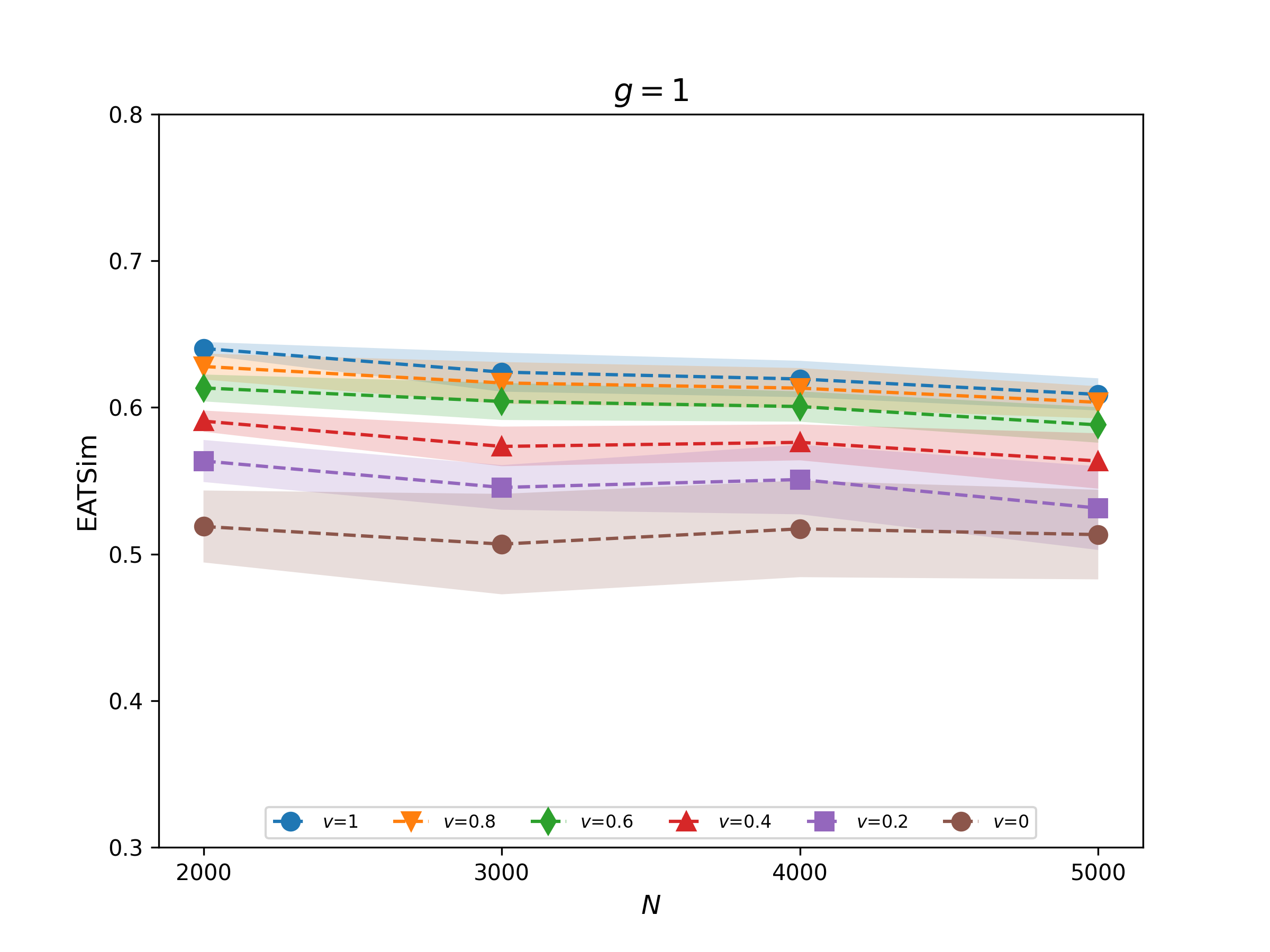}
        % \caption{}
        \label{fig:EATSim_v}
        \begin{picture}(0,0)
            \put(-110,160){\large \textbf{(d)}}
        \end{picture}
    \end{subfigure}
     \caption{\textbf{Quantifying interlayer similarity in synthetic networks with EATSim. }\textbf{(a)} EATSim values between the original network and 19 additional networks with varying rewiring probabilities. \textbf{(b)} Heatmap illustrating EATSim similarity across 20 BA networks, each labeled with its respective rewiring percentage. \textbf{(c)} EATSim shows a strong correlation with angular similarity in synthetic interconnected networks generated with GMM, with higher angular similarity corresponding to larger EATSim values and more similar topologies. \textbf{(d)} Similar to \textbf{(c)}, but for varying radial correlation values with fixed angular correlation. }
    \label{fig:syn}
\end{figure}
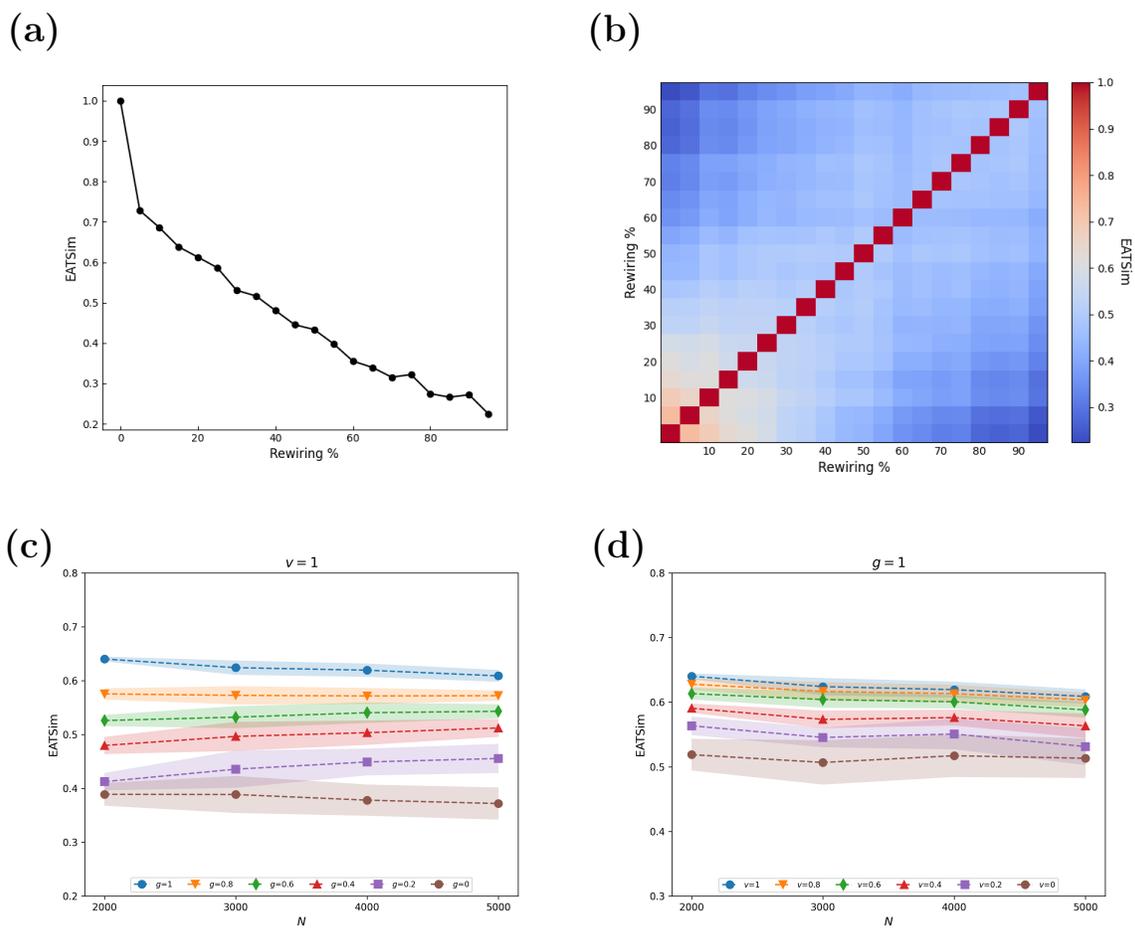

The rewiring probability serves as a benchmark for assessing EATSim's effectiveness in quantifying interlayer similarity. Figure~\ref{fig:syn}\textbf{(a)} illustrates a monotonic decrease in EATSim values as the rewiring probabilities increase between the original and subsequent BA networks, demonstrating the sensitivity of EATSim to varying levels of similarity. Figure~\ref{fig:syn}\textbf{(b)} depicts the relationship between EATSim values and pairs of BA networks with different rewiring probabilities. Networks with lower rewiring probabilities show greater structural similarity and higher EATSim scores, confirming the accuracy of EATSim in quantifying interlayer similarity.

In addition to generating interdependent networks with varying similarities by controlling the link rewiring probabilities of BA networks, we also employ the Geometric Multiplex Model (GMM)~\cite{Kleineberg2016Hidden} to produce synthetic multiplex networks with diverse interlayer similarities. Specifically, we adjust the angular correlation parameter \(g\) and the radial correlation parameter \(v\) (both ranging from 0 for no correlation to 1 for maximum correlation).
By modifying \(g\) and \(v\), we can directly control the structural correlations between layers, generating networks with controllable interlayer similarity.

In this study, we generate various networks with GMM by fixing the average node degree at 6, using a power law degree distribution of \(\gamma=2.5\) and a temperature of \(T=0.4\), with network sizes ranging from 2,000 to 5,000. We examine the relationship between EATSim and angular interlayer similarity by fixing the radial correlation at \(v=1\) and varying \(g\) from 0 to 1. As shown in Fig.~\ref{fig:syn}\textbf{(c)}, when the radial correlation is fixed (\(v=1\)), EATSim and the angular correlation \(g\) exhibit a strong positive correlation across all network scales, with higher angular similarity corresponding to larger EATSim values. Similarly, Fig.~\ref{fig:syn}\textbf{(d)} demonstrates that when the angular correlation \(g\) is fixed (\(g=1\)), EATSim effectively captures variations in radial correlation, with higher radial similarity resulting in larger EATSim values. To provide a more comprehensive view of how EATSim responds to the joint effects of angular and radial correlations, Supplementary Fig.~2~\cite{SI} presents a heatmap of the EATSim values over a range of \(g\) and \(v\). We also compare EATSim with four existing methods (detailed descriptions provided in Section~\ref{sec:3_2} ) in measuring angular ($g$) and radial ($v$) correlations (see Supplementary Figs.~7--10~\cite{SI}).
Additionally, the temperature $T$ of the GMM model controls the clustering within the network. The mesoscopic structural changes in communities caused by this parameter can also be captured by our EATSim (see Supplementary Fig.~3~\cite{SI}). 
These findings confirm that EATSim is an effective metric for measuring structural and geometric similarity in synthetic multilayer networks, underscoring its potential for applications such as network robustness prediction and network reduction in real-world multiplex networks.

\subsection{Predicting the robustness of multiplex networks with EATSim.}
\label{sec:3_2}

Multiplex networks are highly susceptible to targeted attacks, where the failure of just a few critical nodes can trigger a cascading collapse that ultimately fragments the entire system~\cite{Buldyrev2010Catastrophic}. Accurately predicting the robustness of multiplex networks is essential for designing resilient and fail-safe systems that can withstand unforeseen disruptions. Such predictions are crucial for the effective management of complex systems, ensuring their reliability and robustness~\cite{Xu2023Interconnectedness}. In this section, we validate that EATSim, which comprehensively encodes both radial and angular correlations, provides a more accurate assessment of network robustness under targeted attacks compared with other network robustness assessment algorithms. 

The Giant Mutually Connected Component (GMCC) refers to the largest subset of nodes that remain mutually connected across all layers of a multiplex network. As nodes or links fail, the size of the GMCC diminishes, reflecting a loss of network connectivity and a degradation of system functionality. Following the method proposed in~\cite{Kleineberg2017Geometric}, we quantify the robustness of interconnected networks by calculating the critical number of nodes, $\Delta N$. The removal of $\Delta N$ nodes reduces the size of the GMCC from more than $\alpha M$ to less than $M^\beta$, where $M$ is the initial size of the GMCC. In this paper, we set $\alpha=0.4$ and $\beta=0.5$. 
Supplementary Fig.~5~\cite{SI} shows that as $\Delta N$ is a metric dependent on network size, networks with a larger number of nodes tend to have larger $\Delta N$. 
% }
To eliminate the influence of network size and make a fair comparison between different multilayer networks, we compare the robustness of interconnected networks with that of their reshuffled counterpart networks, generated by randomly reshuffling the node mapping relationship of the original interconnected networks, where $\Delta N_{\mathrm{r s}}$ represents the critical number of nodes needed for the reshuffled networks. 
The robustness of interdependent networks can be measured with $\Omega$, computed via Eq.~(\ref{eq:Omega}), which quantifies the relative robustness of an interdependent network compared with its reshuffled counterpart with the same size. 
Supplementary Fig.~6~\cite{SI} shows that $\Omega$ is a network size-independent variable, making it a reliable metric for comparing the robustness of different multiplex networks, regardless of their network sizes. Multiplex networks exhibit significant vulnerability to targeted attacks \cite{Kleineberg2017Geometric}, as the removal of several key nodes can lead to the failure of the entire network. In this paper, for simplicity, we consider vital node attack as the cascading failure dynamics to remove nodes according to their maximum degree of both layers, $K_i=max(k_i^{(G^{(\alpha)})},k_i^{(G^{(\beta)})})$, where $k_i^{(G^{(\alpha)})}$ denotes the degree of node $i$ in network $G^{(\alpha)}$. We then sort nodes based on $K_i$ and remove the node with the highest $K_i$, after which we re-evaluate all $K_i$ for the next attack. The sorting and removal process continues until the GMCC is less than the benchmark value of $M^{0.5}$.
\begin{equation}
    \label{eq:Omega}
    \Omega=\frac{\Delta N-\Delta N_{\mathrm{r s}}} {\Delta N+\Delta N_{\mathrm{r s}}}.
\end{equation}

To further validate the effectiveness of EATSim in predicting network robustness, we compare the robustness estimation capability of EATSim with four network similarity measurement algorithms: the Jensen-Shannon divergence-based measure (JSD)~\cite{DeDomenico2015Structural}, the normalized mutual information metric (NMI)~\cite{Kleineberg2017Geometric}, the node-distance probability distribution metric (D-measure)~\cite{Schieber2017Quantification}, and the connecting-edge similarity measure (LSim)~\cite{Zhang2020Measuring}.

\begin{figure}[htbp]
    \centering
    \begin{subfigure}[b]{0.42\textwidth}
        \centering
        \includegraphics[width=\textwidth]{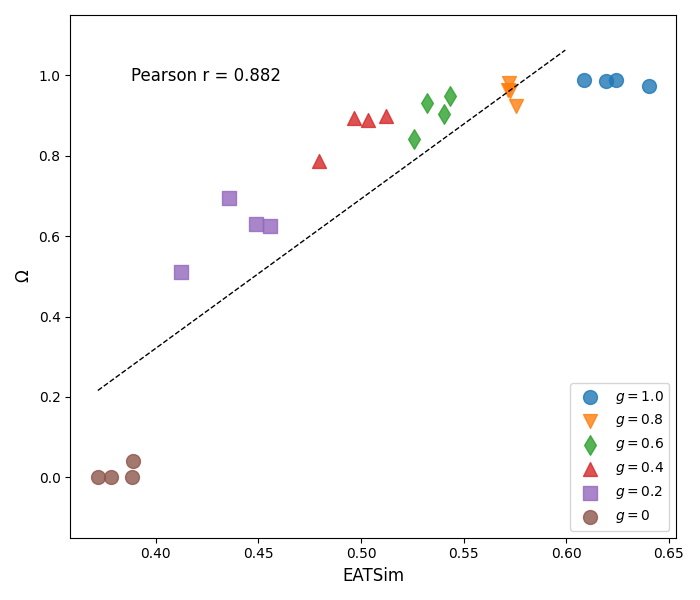}
        % \caption{}
        \label{fig:EATSim_syn}
        \begin{picture}(0,0)
            \put(-110,170){\large \textbf{(a)}}
        \end{picture}
    \end{subfigure}
    \hspace{0.5cm} % 调整这个值来设置间距
    \begin{subfigure}[b]{0.48\textwidth}
        \centering
        \includegraphics[width=\textwidth]{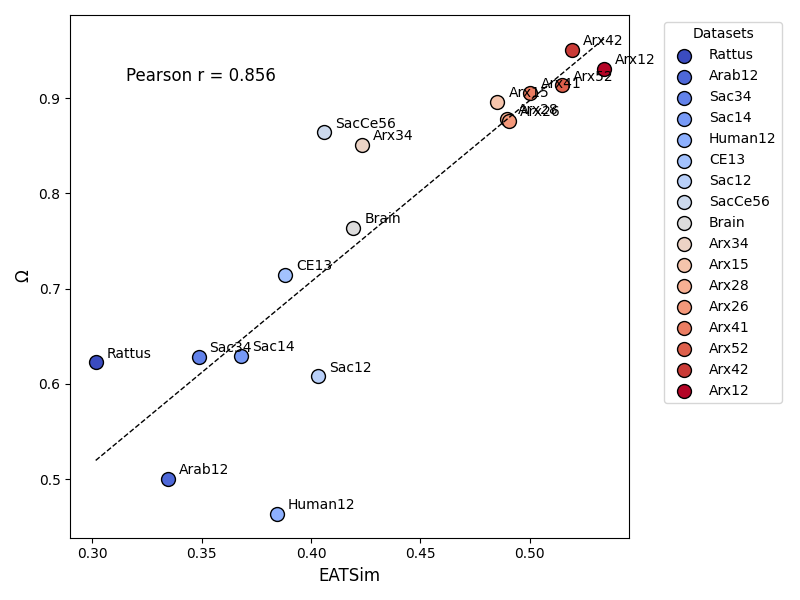}
        % \caption{}
        \label{fig:EATSim_real}
        \begin{picture}(0,0)
            \put(-128,170){\large \textbf{(b)}}
        \end{picture}
    \end{subfigure}
    \caption{\textbf{Robustness $\Omega$ of multiplexes as a function of their EATSim values.} \textbf{(a)} The correlation between robustness and EATSim for synthetic multiplex networks generated by the GMM, with differing sizes ranging from 2,000 to 5,000 for each interlayer angular correlation $g$. \textbf{(b)} The correlation between robustness and EATSim for 17 selected real-world multiplex networks.}
    \label{fig:robustness}
\end{figure}
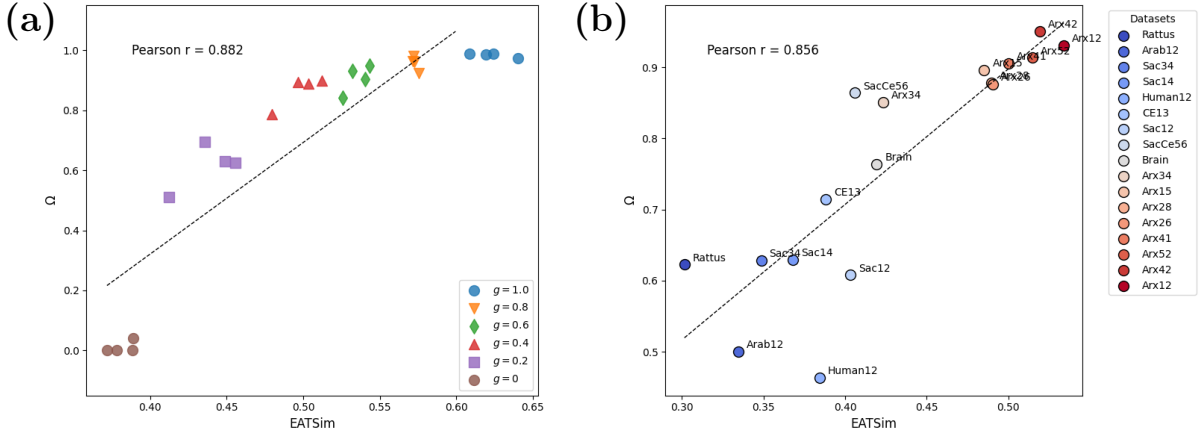

We start by calculating layer similarity using EATSim, alongside the previously mentioned baseline algorithms, on synthetic multilayer networks generated by GMM, varying the angular correlation \( g \) while holding other parameters constant. Subsequently, we apply $\Omega$ to assess the robustness of these synthetic multiplex networks.
As shown in Fig.~\ref{fig:robustness}\textbf{(a)}, networks with higher EATSim scores display higher $\Omega$ values and greater robustness against targeted attacks. The Pearson Correlation Coefficient (PCC) between EATSim and $\Omega$ equals 0.882, higher than other network similarity measurements (see Supplementary Fig.~11 and Table~1~\cite{SI} for more details).

While previous studies reveal that the robustness of multiplex networks depends primarily on angular correlations~\cite{Kleineberg2017Geometric,Wu2021Robustness}, other research shows that radial correlations (i.e., degree correlations) and community overlap also contribute to network stability~\cite{Faqeeh2018Characterizing,Reis2014Avoiding}. Our findings from synthetic multilayer networks generated by the GMM model demonstrate that EATSim can simultaneously capture both angular and radial interlayer correlations, see Fig.~\ref{fig:syn}\textbf{(c)} and \textbf{(d)}. EATSim provides a more comprehensive evaluation of how angular and radial correlation influence network robustness, offering a more reliable metric for assessing the resilience of interconnected systems to targeted attacks.

To extend our analysis, we also validate EATSim with real-world multiplex networks. We select 17 multiplex networks from various domains, such as sociology, biology, and technology.  Figure~\ref{fig:robustness}\textbf{(b)} shows the strong positive correlation between EATSim and $\Omega$ over all networks, with the PCC value equal to 0.856. As shown in Supplementary Fig.~12 and Table~1~\cite{SI}, JSD and D-measure are negatively correlated with $\Omega$, with PCC values equal to -0.333 and -0.648, respectively. Meanwhile, LSim and NMI exhibit positive correlations with $\Omega$, with PCC values of 0.666 and 0.819, respectively. Among all comparison metrics, EATSim shows the highest correlation with $\Omega$, underscoring its superior potential in predicting the robustness of real-world multiplex networks. Furthermore, we calculate the clustering coefficients of these real networks and observe strong positive correlations with both EATSim and $\Omega$ (see Supplementary Fig.~4~\cite{SI}), suggesting that real networks with higher clustering tend to exhibit greater robustness, which is consistent with our findings from the GMM-based simulations.

% \subsection{Predicting the reduction of multiplexes with EATSim.} 
\subsection{Predicting the reducibility of multiplex networks with EATSim}
As the number of layers increases, the computational complexity of structural characteristics for multiplex networks scales superlinearly, while some interactions between layers may be redundant. This redundancy introduces unnecessary complexity without contributing meaningful information and increases the variance in interlayer interactions, rendering the system functionality more fragile to minor perturbations \cite{Kleineberg2017Geometric,Battiston2014Structural}. Network reduction, consolidating the most relevant relationships and creating a clearer and more accurate representation of the structure of complex systems, is a critical requirement when addressing the challenge posed by multiplex systems \cite{DeDomenico2015Structural}. Network reduction can reduce the number of layers in interdependent networks while maximizing the distinguishability between layers by aggregating layers with similar structures. 

Inspired by the Von Neumann entropy and the work of De Domenico et al. \cite{DeDomenico2015Structural}, we quantify the layer difference by assuming that each layer represents a possible network state and apply metric $q$ (see Eq.~(\ref{eq:q})) to measure the discernibility between the multilayer network and the aggregated network. Given a multilayer network $G=\{G^{(\alpha)}\}$, where $\alpha=1,2,...,L$, a particular case is represented by the aggregated graph $A$ associated to $G$, which is the one-layer network whose adjacency matrix is obtained by summing the adjacency matrices of all the $L$ layers of $G$. The Von Neumann entropy of the aggregated graph $A$ is $h_A$. Typically, if we aggregate some of the original layers of $G$, we obtain a reduced multilayer network $C=\{C^{(\alpha)}\}$, where $\alpha=1,2,...,m$ and $m \leq L$ layers. We then consider the entropy per layer of the multilayer network $C$:

\begin{equation}
    \bar{H}(C) = \frac{H(C)}{m} = \frac{\sum_{\alpha=1}^{m} h_{C^{(\alpha)}}}{m},
\end{equation}
and the distinguishability between the multilayer network $C$ and the corresponding aggregated graph $A$ is then quantified through the relative entropy:

\begin{equation}
    \label{eq:q}
    q(C) = 1 - \frac{\bar{H}(C)}{h_{A}}.
\end{equation}

In general, $q$ can either increase or decrease as a result of the aggregation of two layers, depending on several factors such as the relative density of the two graphs or their actual wiring patterns. The optimal configuration of a multilayer network is thus achieved by maximizing the distinction between its layers, which corresponds to maximizing $q$: the larger the value of $q$, the greater the distinguishability between layers and the better the network reduction performance. Intuitively, if the aggregation of two separated layers does not lead to a decrease in $q$, then a reduced configuration, which is more compact, is preferable. 

Identifying similar layers to aggregate is a critical step in network reduction task. Given a multilayer network with $L$ layers, at each reduction step, we first find two layers with the most similar structure and then aggregate them, forming a new multilayer network with one layer less, for which we compute the corresponding value of $q$. Here, we find that EATSim can accurately identify the most similar layers with the highest EATSim value and achieve state-of-the-art network reduction performance. We validate the superior performance of EATSim by comparing it with several other network similarity algorithms on both synthetic and real-world networks.

To test the effectiveness of EATSim, we use nine different real-world networks, encompassing five genetic networks (Candida, Gallus, Plasmodium, Bos, and Human-Herpes4 networks), two social networks (CKM-Physicians-Innovation and CS-Aarhus networks), one transport network (London Tube network), and one neuronal network (C. Elegans multiplex connectome). All networks are obtained from \href{https://manliodedomenico.com/data.php}{Manlio's homepage}. Genetic multiplex networks consist of multiple layers, each representing different types of genetic interactions or relationships. The complexity and redundancy across these layers can obscure key biological insights, making it difficult to analyze and interpret the network as a whole.
Here, we summarize the results obtained by applying the layer aggregation procedure to four genetic multilayer networks, using EATSim and four algorithms mentioned in robustness prediction.

The result of network reduction is a dendrogram (see Supplementary Figs.~14--22~\cite{SI}), that is, a hierarchical clustering diagram that indicates the aggregation procedure of layers. The cut of the dendrogram corresponding to the maximal value of $q$ identifies the optimal configuration of the multilayer network (see Supplementary Tables~3--11~\cite{SI} for details).

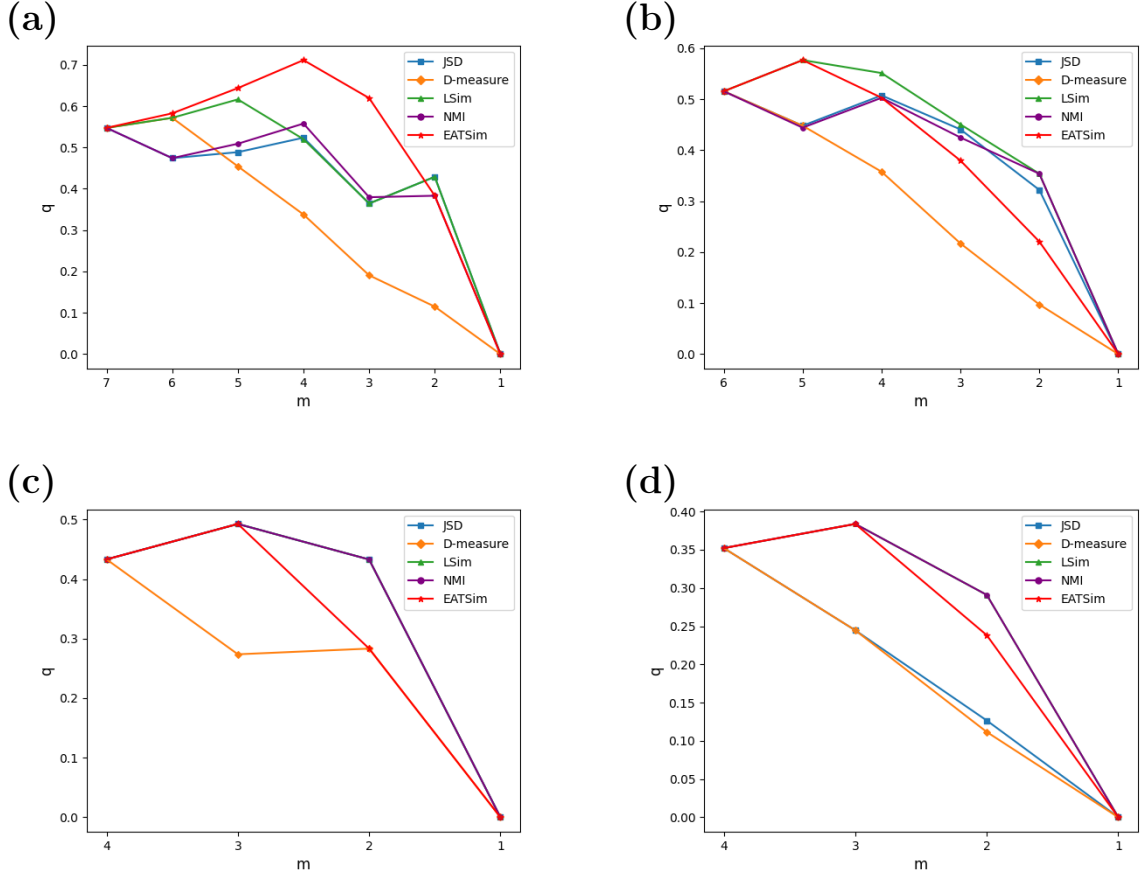
\begin{figure}[htbp]
    \centering
    \begin{subfigure}[b]{0.45\textwidth}
        \centering
        \includegraphics[width=\textwidth]{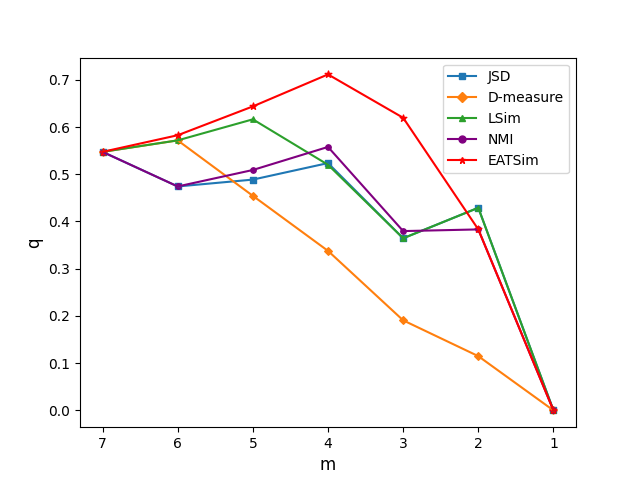}
        % \caption{}
        \label{fig:Candida}
        \begin{picture}(0,0)
            \put(-110,160){\large \textbf{(a)}}
        \end{picture}
    \end{subfigure}
    \hspace{0.5cm} % 调整这个值来设置间距
    \begin{subfigure}[b]{0.45\textwidth}
        \centering
        \includegraphics[width=\textwidth]{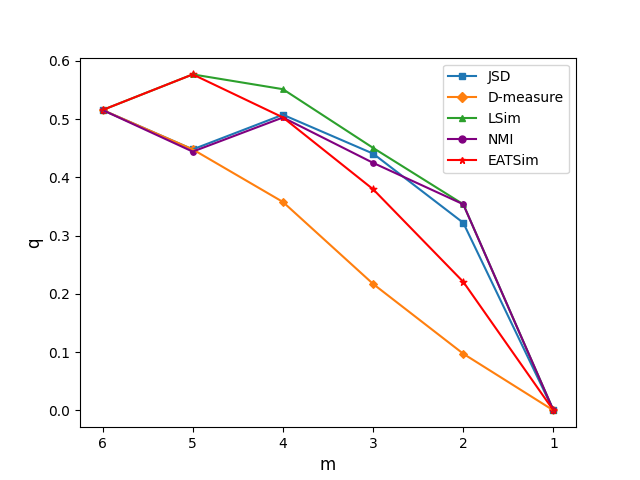}
        % \caption{}
        \label{fig:Gallus}
        \begin{picture}(0,0)
            \put(-110,160){\large \textbf{(b)}}
        \end{picture}
    \end{subfigure}

    \vspace{0.05cm}

    \begin{subfigure}[b]{0.45\textwidth}
        \centering
        \includegraphics[width=\textwidth]{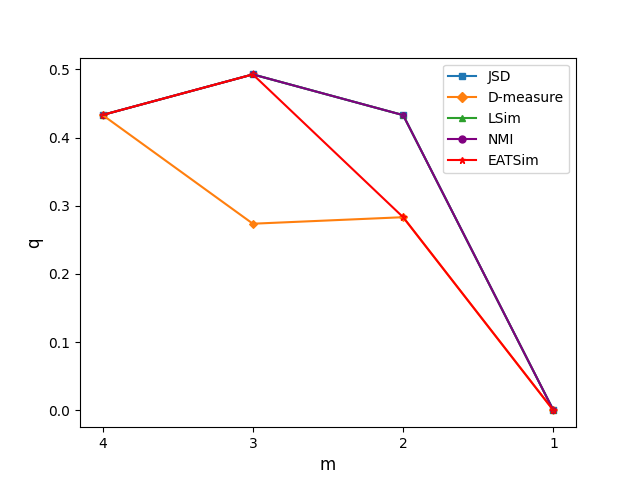}
        % \caption{}
        \label{fig:Bos}
        \begin{picture}(0,0)
            \put(-110,160){\large \textbf{(c)}}
        \end{picture}
    \end{subfigure}
    \hspace{0.5cm} % 调整这个值来设置间距
    \begin{subfigure}[b]{0.45\textwidth}
        \centering
        \includegraphics[width=\textwidth]{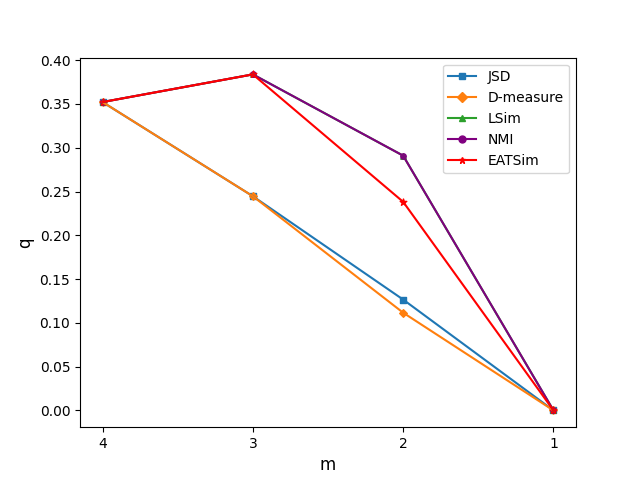}
        % \caption{}
        \label{fig:Human}
        \begin{picture}(0,0)
            \put(-110,160){\large \textbf{(d)}}
        \end{picture}
    \end{subfigure}
    \caption{\textbf{Network reducibility of four genetic networks evaluated using the distinguishability metric \( q \).} \textbf{(a)} In the Candida network, which consists of seven layers, reducing the layer number \(m\) from seven to four enables EATSim to achieve the highest \( q \) value of 0.711, outperforming the other four methods. \textbf{(b)} In the Gallus network, which comprises six layers, EATSim is one of the methods that increases the \( q \) value, reaching a maximum of 0.577 after a single aggregation step. \textbf{(c)} and \textbf{(d)} For the Bos and Human-Herpes4 networks, each with four layers, EATSim attains the maximum \( q \) values of 0.493 and 0.384, respectively, both achieved after the first aggregation step.}
    \label{fig:real_reduction}
\end{figure}

For the Candida network, as shown in Fig.~\ref{fig:real_reduction}\textbf{(a)}, EATSim significantly outperforms the other four algorithms and achieves the maximum value of $q$, equal to 0.711, when the layer number $m$ is reduced from seven to four. For the Gallus, Bos, and Human-Herpes4 networks, as shown in Fig.~\ref{fig:real_reduction}\textbf{(b--d)}, EATSim remains one of the best-performing methods, reaching the maximum $q$ value after the first aggregation step (see Supplementary Table~2~\cite{SI} for details).

Besides, EATSim also performs well on other datasets (see Supplementary Fig.~13~\cite{SI}). Specifically, in the CKM-Physicians-Innovation, C. Elegans, and Plasmodium networks, all five methods achieve comparable performance, reaching the maximum $q$ value when all three layers remain separate. However, EATSim shows an advantage in mitigating the decrease of $q$ value during the reduction of the CKM-Physicians-Innovation network. Notably, during the first layer aggregation, EATSim outperforms the other four methods by selecting the layer pair that results in the smallest decrease in the $q$ value of the multilayer network. In the case of the London multiplex transportation network, while NMI achieves the highest $q$ value of 0.521 when reducing the network from 13 to 3 layers, EATSim closely follows with a $q$ value of 0.487, demonstrating its competitive performance.

\begin{figure}[htbp]
    \centering
    % Left image taking 60% of the width, vertically centered
    % Right column with two images
    \begin{minipage}[b]{0.25\textwidth}
        \centering
        \includegraphics[width=\textwidth]{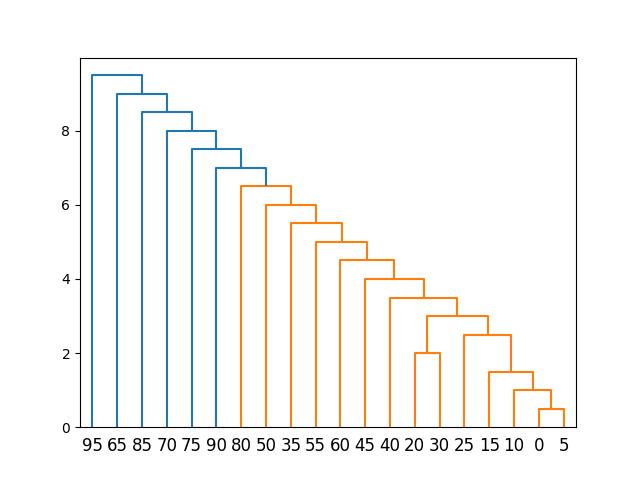}
        \begin{picture}(0,0)
            \put(-70,100){\large \textbf{(a)}}
        \end{picture}
        % \vspace{1em} % Adjust space between the two images in the right column
        \includegraphics[width=\textwidth]{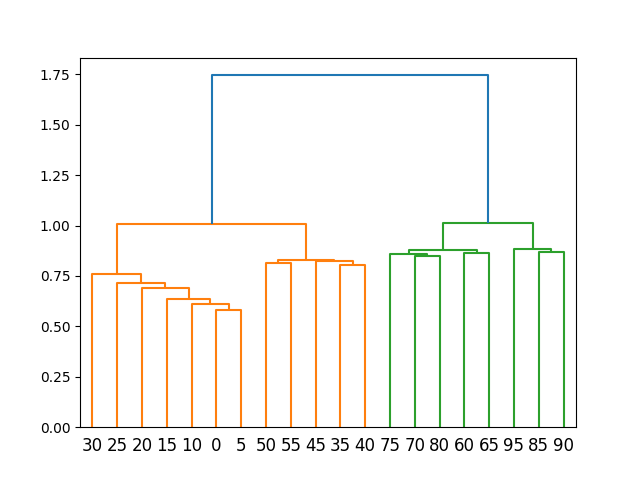}
        \begin{picture}(0,0)
            \put(-70,100){\large \textbf{(b)}}
        \end{picture}
    \end{minipage}
    \hspace{0.5cm} % Adjust the space between the left and right images
    \begin{minipage}[b]{0.55\textwidth}
        % \raisebox{0.2\height}{ % Adjust this value to vertically center the left image
        %     \includegraphics[width=\textwidth]{figures/f_q_syn_reduction.png}
        % }
        \includegraphics[width=\textwidth]{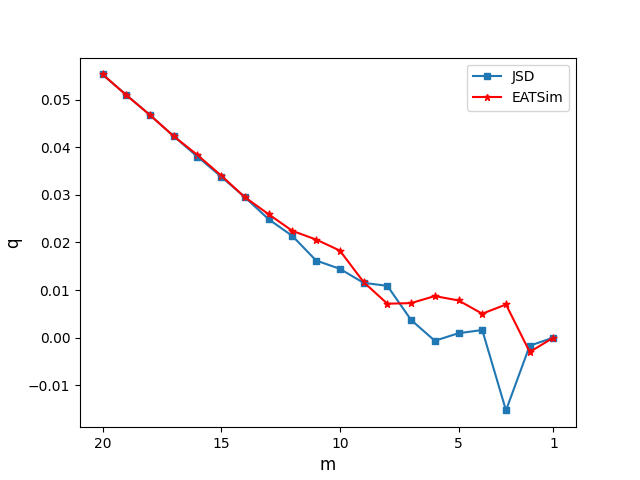}
        \begin{picture}(0,0)
            \put(-10,205){\large \textbf{(c)}}
        \end{picture}
    \end{minipage}
    \caption{\textbf{Network reduction process on synthetic multilayer network.} \textbf{(a)} EATSim can capture the similarity between layers with relatively large differences in rewiring rates, , rather than always aggregating layers with adjacent rewiring rates. \textbf{(b)} The hierarchical clustering procedure of JSD-based network reduction merges layers with adjacent rewiring rates preferentially. \textbf{(c)} The $q$ function is decreasing throughout the entire reduction process of both JSD and our EATSim, with the network reaching its optimal state when no layers are aggregated.}
    \label{fig:syn_reduction}
\end{figure}
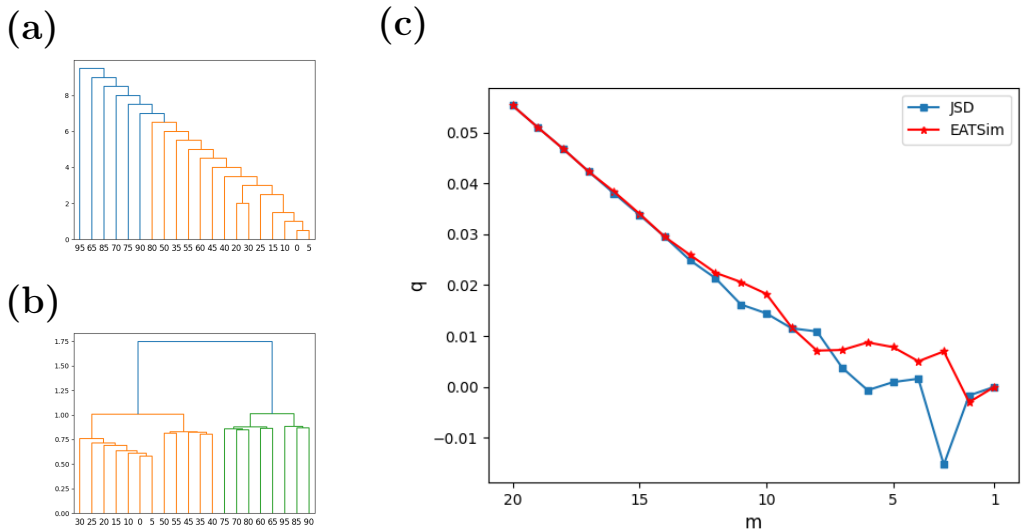

We also conduct experiments with the aforementioned synthetic networks in Section~\ref{sec:3_1}, which are generated by controlling the link rewiring probabilities of BA networks, with probabilities ranging from 5\% to 95\%. 
As shown in Fig.~\ref{fig:syn_reduction}\textbf{(a)}, the hierarchical clustering procedure first aggregates similar layers characterized by small rewiring probabilities, and then proceeds to the aggregation of dissimilar layers obtained by large rewiring probabilities. The results suggest that layers with high edge overlap and similar structure tend to be aggregated earlier, and this trend is similar to the reduction result generated by Jensen-Shannon divergence (JSD) \cite{DeDomenico2015Structural}, see Fig.~\ref{fig:syn_reduction}\textbf{(b)}. 
However, unlike JSD, EATSim-based network reduction does not always aggregate two layers with adjacent rewiring probabilities. It sometimes finds similarities between layers with larger rewiring differences, which JSD fails to capture. This capacity of EATSim is demonstrated to be effective in the result of $q$, shown in Fig.~\ref{fig:syn_reduction}\textbf{(c)}. Although the best network configuration in this case is the initial one without layer aggregation, EATSim, compared with JSD, alleviates the decrease in the $q$ value of the network to a greater extent.

\section{Conclusion}

In this paper, we address the problem of quantifying interlayer similarity of multiplex networks from the perspectives of network representation and geometric learning. We propose a novel similarity measurement algorithm, EATSim, which combines the PED loss and AED loss to capture both local and global structural characteristics of networks. EATSim demonstrates remarkable generalization capabilities across diverse synthetic and real networks. To validate its effectiveness, we evaluate EATSim on network robustness and reducibility prediction tasks, highlighting its great advantage in capturing hidden structural properties in multiplex networks. Furthermore, EATSim serves as a reliable predictor of network robustness against attacks and achieves improved reduction results for multilayer networks.

The successful utilization of EATSim in assessing multiplex network robustness and reducibility underscores its broad applicability and potential in designing effective network protection strategies and obtaining a compact and informative configuration of interdependent multiplexes. Our findings suggest that EATSim is a valuable tool for analyzing complex network structures and holds promise for future research and applications.

\section*{Author contributions}

\textbf{H.N.} and \textbf{S.W.} contributed equally to this work. \textbf{H.N.} and \textbf{S.W.} contributed Methodology, Experiments, Analysis, and Writing; \textbf{C.O., Y.Z.} contributed Data Collection, Experiments; \textbf{W.G.} contributed Supervision, Methodology, Conceptualization, Writing, and Funding Acquisition. 

\section*{Declaration of Competing Interest}

The authors declare that they have no known competing financial interests or personal relationships that could have appeared to influence the work reported in this paper.

\section*{Acknowledgements} 

This work was supported by grants from the National Natural Science Foundation of China (Grant No.42450183) and the Beijing University of Chemical Technology (Grant No. PY2514, ZY2412 and 11170044127). 

\noindent
\bibliographystyle{unsrt}
\bibliography{main}
% \end{thebibliography}

\end{document}